\documentclass[showpacs,twocolumn,showkeys,amsmath,amssymb,pra]{revtex4-1}

\usepackage{graphicx}   % Include figure files

\begin{document}

\title{Cavity resonances dominating the photon statistics in the non-equilibrium steady state} 

\author{Felix R\"uting}
\email{felix.ruting@uam.es}
\affiliation{Departamento de F{\' i}sica Te\'orica de la Materia Condensada, Universidad Aut{\'o}noma de Madrid, Madrid 28049, Spain} 
\affiliation{Institut f\"ur Physik, Carl von Ossietzky Universit\"at, D-26111 Oldenburg, Germany}

\author{Christoph Weiss}
\email{christoph.weiss@durham.ac.uk}
\affiliation{Joint Quantum Centre (JQC) Durham--Newcastle, Department of Physics, Durham University, Durham DH1 3LE, United Kingdom}
\affiliation{Institut f\"ur Physik, Carl von Ossietzky Universit\"at, D-26111 Oldenburg, Germany }


%\keywords{}
                  
\date{\today}
 
\begin{abstract}
The non-equilibrium-steady state (NEST) for photons in a cavity is
investigated theoretically. The NEST is caused by different parts of
the cavity being at distinct temperatures or by temperature gradients.
By using a rate equation based on the Lindblad equation, we derive an
analytic expression for the steady-state distribution of the photon
spectrum. We predict differences between the non-equilibrium steady state
and a fit to the black-body spectrum calculated via Planck's law with
an effective temperature. For two bodies of similar size at two
temperatures which differ by a factor of two, the difference would be
more than 10\%. We also show that cavity resonances have a particularly
large influence on the resulting non-equilibrium steady state of the
photons.  The investigation of thermal spectra in the presence of more
than one temperature can be important for high-precision atomic
clocks.
\end{abstract} 
\maketitle 

%%%%%%%%%%%%%%%%%%%%%%%%%%%%%%%%%%%%%%%%%%%%%%%%%%%%%%%%%%%%%%%%%%%%%%%%%%%%%%%%

\section{\label{sec:introduction}Introduction}

Planck's law describes radiation from a black body in a cavity in
thermal equilibrium~\cite{LandauLifshitz80}. It is an important
contribution to the development of quantum mechanics; the law plays a
role in fields as diverge as photonic crystals~\cite{LinEtAl98} and
cosmology~\cite{RollWilkinson66,PeeblesRatra03}. The cavity involved
in derivations of Planck's law is used in many quantum optics
experiments (see, e.g., Refs.~\cite{RauschenbeutelEtAl99,
  LounisEtAl00, PinskeEtAl00,EnglundEtAl07} and references therein).
Theoretical investigations include quenching and spontaneous
emission~\cite{LeeEtAl97}, entanglement distribution among distant
nodes in a quantum network~\cite{CiracEtAl97}, quantum optics with
surface plasmons~\cite{ChangEtAl06} and quantum phase transitions of
light~\cite{GreentreeEtAl06}.

The focus of the present paper lies on the photon-distribution for
cases where more than one temperature is involved. For such
situations, the non-equilibrium steady state, NEST, is
investigated. Such a distribution could be measured experimentally in
state-of-the-art quantum optics experiments. However, its applications
would not be restricted to quantum optics.

Cavities bordered by bodies at different temperatures appear quite
natural, e.g., by studying the temperature dependence of the
Casimir-interaction~\cite{Lifshitz_56,Haakh_09,Haakh_10,Antezza_06,
  Antezza_08, Mohideen_98, Decca_03, Decca_05, Obrecht_07} or the
near-field radiative heat transfer~\cite{Perez_09, Chapuis_08,
  Dorofeyev_07, Pendry_99, Hu_08, Kittel_08, Wischnath_08,
  Narayanaswamy_08, Rousseau_09, Shen_09,
  Altfeder_10,BenAbdallahBiehs2014}. Indeed, this transfer between
bodies separated by distances even below $1\,\text{nm}$ and with
temperature differences up to some $100\,\text{K}$ have been
investigated in several experiments~\cite{Hu_08, Kittel_08,
  Wischnath_08, Narayanaswamy_08, Rousseau_09, Shen_09,
  Altfeder_10,Ottens2011}, recently. Altfeder \emph{et al.}, for
example, have studied the heat transfer between a STM tip and a sample
with a separation of only a few angstrom and with a maximal
temperature difference of about
$200\,\text{K}$~\cite{Altfeder_10}. Therefore a theoretical
description of these experiments requires the understanding of the
photon field in non-equilibrium states and different geometries.

Another important near-field effect is the Casimir force. For metal
bodies in thermal equilibrium separated by a distance well below the
thermal wavelength the effect of finite temperature is only a relative
small correction~\cite{Lifshitz_56}, but recent progress in the
measurement of this force allows one to study even this correction
both for two bulk bodies and for a bulk body and a gas-phase
atom~\cite{Mohideen_98, Decca_03, Decca_05,Obrecht_07}. The Casimir
force was also investigated theoretically for two parallel plates at
different temperatures~\cite{Antezza_06,Antezza_08}. However, these
studies are limited to the case of two parallel plates since the
calculation requires the knowledge of the fluctuating fields between
the bodies and for the case of a plane cavity these fields were
calculated by Dorofeyev \emph{et al.}~\cite{Dorofeyev_02}.

Finally, for optical atomic
clocks~\cite{DiddamsEtAl2001,TakamotoEtAl2005,RosenbandEtAl2008,BloomEtAl2014},
high-precision experiments have reached a regime where the accuracy
can be limited by the black-body
radiation~\cite{MiddelmannEtAl2012,BloomEtAl2014,HinkleyEtAl2013}. Recently, accuracy
and stability on the 10$^{-18}$ level was reported in an optical
lattice clock for which a temperature gradient near the
lattice-confined atoms was observed~\cite{BloomEtAl2014}. Thus,
calculating deviations from the black-body spectrum due to the
influence of more than one temperature can be important for
high-precision atomic clocks.

Thus, knowing the photon statistic in a NEST is an important step for
determine other field-related quantities. For example, the local
energy density is given by product of the local density of
states~\cite{Joulain_03} and the mean energy of the mode in the actual
state of the system.

The paper is organized as follows: In Sec.~\ref{sec:lindblad} we
introduce the Lindblad equation. Section~\ref{sec:beyond} solves this
master equation for two temperatures as has been realized, e.g.\ for
two parallel plates at distinct temperatures; these results can easily
be generalized to the case of more than two
temperatures. Section~\ref{sec:energy} shows that there are
significant deviations of the non-equilibrium steady state from the
equilibrium distribution.  The paper ends with a conclusion in
Sec.~\ref{sec:conclusion}.

\section{\label{sec:lindblad}Rate equation}
The Lindblad equation~\cite{HarocheRaimond06,Breuer06} for the cavity
field can be found in the appendix. For the purpose of the present
paper, it is sufficient to use the simplified version. Rather than
having to use the complete density matrix in the Lindblad
equation~\cite{HarocheRaimond06,Breuer06}
[cf.\ Eq.~(\ref{eq:lindblad})], one can use a rate equation for the
probabilities $p_{n}(t)$ to find $n=0,1,2,\ldots$ photons in mode
$\omega$ in the cavity at time $t$:
\begin{eqnarray}
\label{eq:masterp}
\frac {d}{dt}p_n(t) &=& \kappa_-(n+1)p_{n+1}(t) +\kappa_+np_{n-1}(t) \nonumber\\
& &-\left[\kappa_-n+\kappa_+(n+1)\right]p_n(t)\;.
\end{eqnarray}
 The non-negative coefficients $\kappa_{\pm}$ correspond to the
 temperature-dependent rates at which photons are emitted into the
 cavity, $\kappa_+$, or absorbed at its boundaries, $\kappa_-$. The
 temperature dependence can be split into a strong temperature
 dependence included at the beginning of Sec.~\ref{sec:beyond} and an
 additional, material-dependent temperature dependence discussed in
 Sec.~\ref{sec:meaning}. The energy of $n$ photons in the mode
 characterized by the frequency
\begin{equation}
\nu\equiv\omega/(2\pi)
\end{equation}
 is given by
\begin{equation}
\label{eq:Eeigen} 
 E = \hbar\omega\left(n+\frac12\right)
\end{equation}

The derivation of the  non-equilibrium steady state will include the equilibrium distribution 
\begin{equation}
\label{eq:nteq}
\langle n\rangle = \frac1{\exp(\hbar\omega\beta)-1}\;,\quad \beta\equiv\frac1{k_{\rm B}T}
\end{equation}
as a special case if all temperatures are equal to~$T$ ($k_{\rm B}$ is the Boltzmann constant). 

\section{\label{sec:beyond}Non-equilibrium steady state (NEST)}
\subsection{\label{sec:two}Two temperatures}

We start with the situation that there are two distinct temperatures
\begin{equation}
\beta_j =\frac 1{k_{\rm B}T_j}\;,\quad j=1,2\;.
\end{equation}
 involved; at the end of this section, the results will be generalized to the situation with more than two temperatures.
We now take Eq.~(\ref{eq:masterp}) and replace
\begin{equation}
\kappa_-\rightarrow\sum_{j=1}^2 \kappa_{-,j}
\end{equation}
and 
\begin{equation}
\kappa_+\rightarrow\sum_{j=1}^2 \kappa_{+,j}\;.
\end{equation}

The temperatures enter into the master equation via the Boltzmann law (cf.\ Ref.~\cite{HarocheRaimond06}) based on the assumption of local thermal equilibrium in each of the bodies:
\begin{equation}
\kappa_{+,j} = \kappa_{-,j}\exp\left(-\hbar\omega \beta_j\right)\;,\quad j=1,2\;.
\end{equation}

As the next step, we introduce a function which, at the present stage of the derivation has no physical meaning. The physical meaning [partially already indicated by our knowledge of Eq.~(\ref{eq:nteq})] will become clear at the end of our calculation. The function reads: 
\begin{equation}
\label{eq:Fj}
F(\beta)\equiv \frac 1{\exp(\hbar \omega \beta)-1}\;.
\end{equation}
With this definition, we have:
\begin{equation}
\frac{\kappa_{+,j}}{\kappa_{-,j}} = \frac{F(\beta_j)}{1+F(\beta_j)}
\end{equation}
and we can define two rates $\kappa_j$ with:
\begin{equation}
\label{eq:kappaminus}
\kappa_{-,j} = \kappa_j[1+F(\beta_j)]
\end{equation}
and
\begin{equation}
\label{eq:kappaplus}
\kappa_{+,j} = \kappa_jF(\beta_j)
\end{equation}
where 
\begin{equation}
\label{eq:kappaj}
\kappa_j>0
\end{equation} might still be temperature-dependent (cf.\ Sec.~\ref{sec:meaning}).

The condition for the non-equilibrium steady state reads
\begin{equation}
\frac{d}{dt}p_n(t)=0\quad {\rm for~all~} n
\end{equation}
which leads to the two (equivalent) conditions
\begin{equation}
\left(\kappa_{+,1}+\kappa_{+,2}\right)np_{n-1}(t) = \left(\kappa_{-,1}+\kappa_{-,2}\right)np_{n}(t)
\end{equation}
and
\begin{equation}
\left(\kappa_{+,1}+\kappa_{+,2}\right)(n+1)p_{n}(t) = \left(\kappa_{-,1}+\kappa_{-,2}\right)(n+1)p_{n+1}(t).
\end{equation}
Taking any of these equations and using that $\kappa_{-,j}$ is positive [see Eqs.~(\ref{eq:Fj}), (\ref{eq:kappaminus}) and (\ref{eq:kappaj})] while $\kappa_{+,j}$ is positive for positive temperatures [Eqs.~(\ref{eq:Fj}), (\ref{eq:kappaplus}) and (\ref{eq:kappaj})]:
\begin{eqnarray}
p_n(t)&=&bp_{n-1}(t)\\
b&\equiv&\frac{\kappa_{+,1}+\kappa_{+,2}}{\kappa_{-,1}+\kappa_{-,2}}\quad {\rm with}\\
0&< &b< 1\quad {\rm for~} T_1+T_2 >0\;.
\end{eqnarray}
The case that both temperatures are zero can be discarded.

The fact that now~$p_n\propto b^n$
allows to calculate
\begin{eqnarray}
\langle n\rangle &=& \sum np_n(t)\\
&=&\frac{\sum_{n=0}^{\infty}nb^n}{\sum_{n=0}^{\infty}b^n} \nonumber\\
&=&\frac1{\frac1b-1}\;.
\label{eq:einsetzen}
\end{eqnarray}
Using 
\begin{eqnarray}
\frac1b &=& \frac{\kappa_1\left(1+F(\beta_1)\right)+\kappa_2\left(1+F(\beta_2)\right)}{\kappa_1F(\beta_1)+\kappa_2F(\beta_2)}\\
&=& \frac{\kappa_1+\kappa_2}{\kappa_1F(\beta_1)+\kappa_2F(\beta_2)} +1
\end{eqnarray}
which can be inserted in Eq.~(\ref{eq:einsetzen})
\begin{equation}
\label{eq:wichtig1}
\langle n\rangle_{\rm NEST}  =\frac{\kappa_1F(\beta_1)+\kappa_2F(\beta_2)}{\kappa_1+\kappa_2}\;,
\end{equation}
where NEST refers to the non-equilibrium steady state.
A relevant special case is $T_1=T_2=T$ for which because of $F(\beta_1)=F(\beta_2)$ we find the expected result~\cite{HarocheRaimond06}
\begin{equation}
\langle n\rangle_{\rm equil.} =\frac1{\exp\left(\frac{\hbar\omega}{k_{\rm B}T}\right)-1}\;.
\end{equation}
For  $\kappa_1=\kappa_2$ we find: 
\begin{equation}
\label{eq:gleichekappa}
\langle n\rangle_{\rm NEST} =\frac12\left(\frac1{\exp\left(\frac{\hbar\omega}{k_{\rm B}T_1}\right)-1}+
\frac1{\exp\left(\frac{\hbar\omega}{k_{\rm B}T_2}\right)-1}
\right).
\end{equation}
This equation and its more general versions [see
  Eqs.~(\ref{eq:wichtig1}) and (\ref{eq:NESTalg})] is the first main
results of the present paper.

\subsection{\label{sec:several}Several temperatures}
Extend the calculation of Sec.~\ref{sec:two} to the case of several temperatures leads to:
\begin{equation}
\label{eq:NESTalg}
\langle n\rangle_{\rm NEST} =\frac{\sum_j\kappa_jF(\beta_j)}{\sum_j\kappa_j}\;,
\end{equation}
or, in general:
\begin{equation}
\langle n\rangle_{\rm NEST} =\frac{\int d^2 r\kappa(\vec{r})F(\vec{r})}{\int d^2 r\kappa(\vec{r})}\;.
\end{equation}
where the integral extends over the surface surrounding the cavity.

\subsection{\label{sec:meaning}Physical meaning of the constants}
What remains to be done is to find a physical interpretation of the parameters $\kappa_{\pm}$. So far, they seem to be purely phenomenological constants. However, similar to the equilibrium case discuss, e.g., in Ref.~\cite{HarocheRaimond06}, they are related to the time-scales at which the non-equilibrium steady state is reached:
\begin{equation}
\frac{d}{dt}\langle n\rangle(t) = \sum_{n=0}^{\infty}n \frac{d}{dt}p_n(t).
\end{equation}
Using the general expression~(\ref{eq:NESTalg}) combined with the master equation~(\ref{eq:masterp}), we find
\begin{equation}
\label{eq:timescales}
\frac{d}{dt}\langle n\rangle(t) = -\left(\sum_j\kappa_j\right)\left[\langle n\rangle(t) - \langle n\rangle_{\rm NEST}\right]\;.
\end{equation}
For a single $\kappa$, this time-scale can easily be related to a dimensionless mode quality factor~$Q$ of a cavity via $\kappa =\omega/Q$~\cite{HarocheRaimond06}.

Because of material-properties, the constants $\kappa_j$ might be temperature dependent.\\

\section{\label{sec:energy}NEST-equivalent of Planck's law}

Because of Eq.~(\ref{eq:Eeigen}), the  average energy in the mode~$\omega$ is given by
\begin{equation}
\langle E\rangle =\hbar\omega \left(\langle n\rangle_{\rm NEST}+\frac12\right)\;.
\end{equation}
However, the zero point energy will not be relevant in our derivation. If one starts, e.g., with the zero-photon situation and lets the system approach to the NEST, it is clear that no energy is transferred into the vacuum state. For the purpose of our calculation we can thus use:
\begin{equation}
\langle E\rangle =\hbar\omega\langle n\rangle_{\rm NEST}
\end{equation}
While for near-field effects more complicated densities of state are relevant, to derive the NEST-equivalent of the Planck's law~\cite{LandauLifshitz80},
\begin{equation}
\label{eq:planck}
I(\omega,T)=\frac{\hbar\omega^3}{\pi^2c^3}\frac1{\exp(\hbar\omega\beta)-1},
\end{equation}
we use the vacuum density~\cite{LandauLifshitz80,Joulain_03} in the following 
\begin{equation}
\label{eq:DOSfree}
D = \frac{\omega^2}{\pi^2c^3}\;.
\end{equation}
Thus, 
\begin{equation}
\label{eq:nestplanck}
I_{\rm NEST}(\omega,\{T_j\})=\frac{\hbar\omega^3}{\pi^2c^3}\langle n\rangle_{\rm NEST},
\end{equation}
\begin{widetext}
or, for the two-temperature case with $\kappa_1=\kappa_2$:
\begin{equation}
\label{eq:two}
I_{\rm NEST}(\omega,\{T_1,T_2\}) = \frac{\hbar\omega^3}{2\pi^2c^3}\left(\frac1{\exp\left(\frac{\hbar\omega}{k_{\rm B}T_1}\right)-1}+
\frac1{\exp\left(\frac{\hbar\omega}{k_{\rm B}T_2}\right)-1}
\right)
\end{equation}
\end{widetext}

In an experiment, a natural way to approach the non-equilibrium steady state data would be to try and fit the equilibrium distribution~(\ref{eq:planck}) using the temperature as a fitting parameter. 
For the prediction of Eq.~(\ref{eq:two}) to be of practical use, we still have to show  that in an experiment it would be distinguishable from a Planck-distribution with an effective temperature~$T_{\rm eff}$. In the following, we also use:
\begin{equation}
\beta_{\rm eff} = \frac1{k_{\rm B}T_{\rm eff}}\;.
\end{equation}
 A suitable way to define such an effective temperature is to minimize the mean square deviations
\begin{equation}
\label{eq:delta}
\delta \equiv \int_0^{\infty}d\omega\left[I_{\rm NEST}(\omega,\{T_1,T_2\})-I(\omega,T_{\rm eff})\right]^2\;.
\end{equation}
In order to visualize the differences between the NEST-distribution and the equilibrium distribution with an effective temperature, we use:
\begin{equation}
\label{eq:Delta}
\Delta \equiv \frac{I_{\rm NEST}(\omega,\{T_1,T_2\})-I(\omega,T_{\rm eff})}{\max\left\{I(\omega,T_{\rm eff}), 0\le\omega<\infty\right\}}\;.
\end{equation}

 According to Wien's displacement law~\cite{LandauLifshitz80}, $I(\omega,T_{\rm eff})$ reaches its maximum at~\footnote{The analytic expression was obtained by using the computer algebra program Maple.}
\begin{eqnarray}
\hbar\omega_{\rm max} = \left\{3+W_0\left[-3\exp(-3)\right]\right\}{k_{\rm B}T_{\rm eff}}\\
\simeq 2.82 {k_{\rm B}T_{\rm eff}}
\end{eqnarray}
where $W_0$ is a Lambert W function. The value of the maximum is
\begin{eqnarray}
 \frac{\{W_0[-3\exp(-3)]+3\}^3}{\{\exp(W_0[-3\exp(-3)]+3)-1\}}\frac{\left(k_{\rm B}T_{\rm eff}\right)^3}{\pi^2c^3\hbar^2}
\nonumber\\\simeq 1.42\frac{\left(k_{\rm B}T_{\rm eff}\right)^3}{\pi^2c^3\hbar^2}\;.
\end{eqnarray}

\begin{figure}
\includegraphics[width=0.9\linewidth]{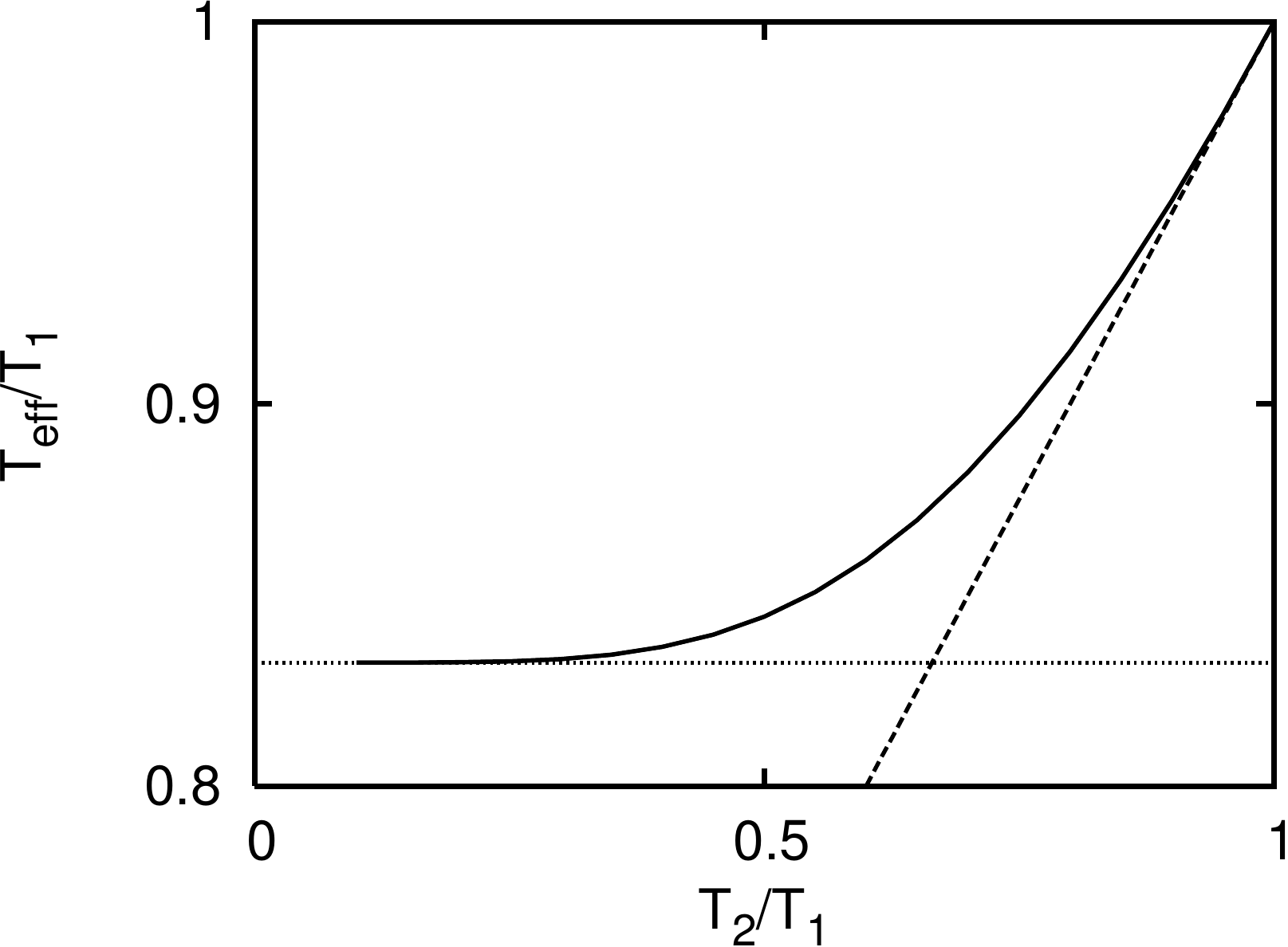}
\caption{\label{fig:Teff}Effective temperature as a function of the ratio of the two temperatures. The effective temperature tries to describe the non-equilibrium steady state by an equilibrium distribution with an effective temperature. Solid line: the effective temperature was obtained by minimizing Eq.~(\ref{eq:delta}). Dotted line: the low temperature limit given in Eq.~(\ref{eq:lowhigh}). Dashed line: the arithmetic mean of the two temperatures $T_1$ and $T_2$ [the high temperature limit of Eq.~(\ref{eq:lowhigh})].}
\end{figure}

Figure~\ref{fig:Teff} displays the effective temperature obtained by minimizing Eq.~(\ref{eq:delta}) as a function of the ratio~$T_1/T_2$.
Without loss of generality we can assume 
\begin{equation}
0<T_2\le T_1<\infty
\end{equation}
for our discussion. Figure~\ref{fig:Teff}
 shows that the effective temperature approaches
\begin{equation}
\label{eq:lowhigh}
T_{\rm eff} \simeq \left\{\begin{array}{lcr}\frac{T_1+T_2}2&:& T_2/T_1\to 1\\ 0.83T_1&:& T_2/T_1\to 0\end{array}\right.\;.
\end{equation}
However, the main point in this approach was not to calculate an effective temperature but rather to see if the NEST-prediction~(\ref{eq:two}) can be distinguished from the effective temperature approach.

\begin{figure}
\includegraphics[width=0.9\linewidth]{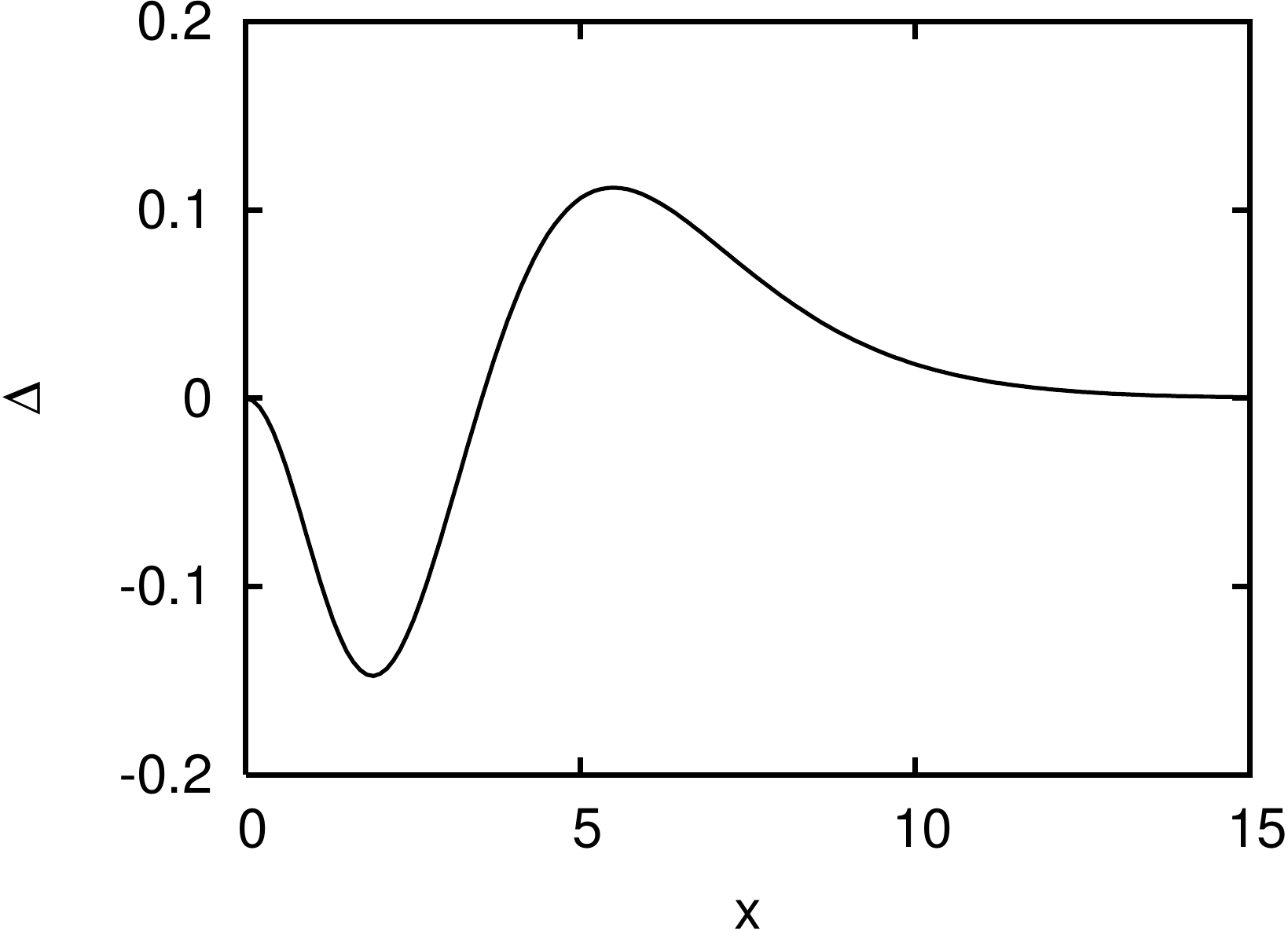}
\includegraphics[width=0.9\linewidth]{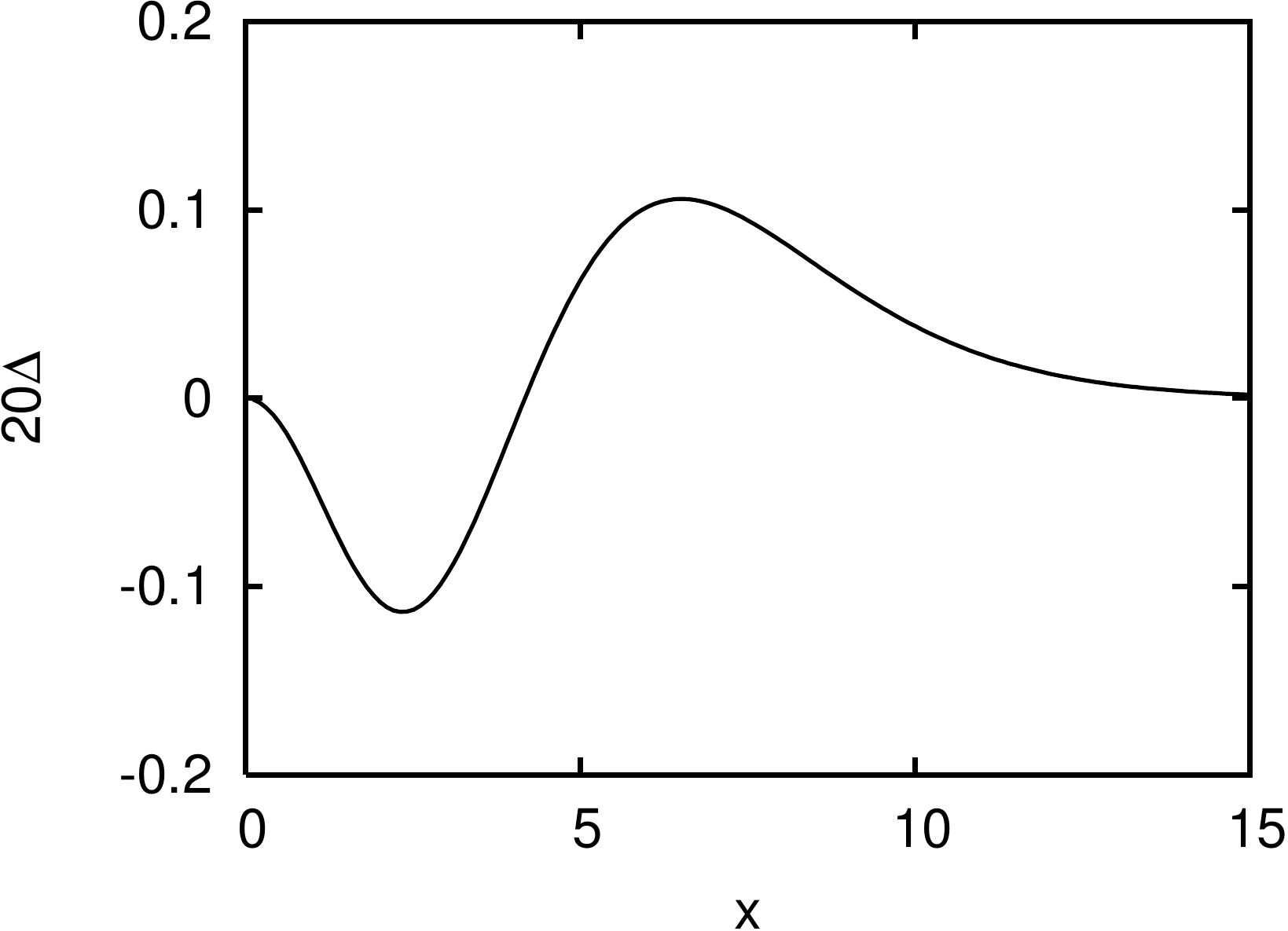}
\caption{\label{fig:Delta}Trying to fit an equilibrium distribution to
  the non-equilibrium steady state with two
  temperatures~(\ref{eq:two}) leads to experimentally detectable
  differences [cf.\ Eq.~(\ref{eq:Delta})]. The difference $\Delta$ is
  plotted as a function of dimensionless
  frequency~$x\equiv\hbar\omega/(k_{\rm B}T_{\rm eff})$ Upper panel:
  $T_2=0.5T_1$, $T_{\rm eff} \simeq 0.8443 T_1$. Lower panel:
  $T_2=0.9T_1$, $T_{\rm eff} \simeq 0.9533T_1$. Note that the error is
  multiplied by 20 in the lower panel. As expected, the deviation of
  the non-equilibrium steady state from thermal equilibrium thus is
  much larger if the ratio $T_2/T_1$ differs from one by an
  experimentally realistic~\cite{Altfeder_10} factor of two (upper
  panel) than if the ratio is close to one (lower panel). }
\end{figure}
Figure~\ref{fig:Delta} shows the deviation of both approaches as
defined in Eq.~(\ref{eq:Delta}). As expected from closely inspecting
Eq.~(\ref{eq:two}), the differences are small if the temperature
difference is small. However, they can be quite large for larger
temperature differences: if the two temperatures differ by a factor of
the order of two (e.g., room-temperature versus liquid nitrogen) the
deviations are more than 10\% (measured in units of the maximum of the
curve) and should thus be easily detectable in an experiment.

\subsection{Cavity}
In order to obtain a NEST-equivalent of Planck's law we assumed in the
previous section that the density of states between the two bodies is
given by the free-space DOS (see. Eq.~(\ref{eq:DOSfree})). However in
an experiment involving a cavity this cavity will strongly influence
the density of states. As a generic example we assume, that the resonance has
a Lorentzian shape~\cite{Garraway96}. Hence in following the density
of states is given by
\begin{equation}
\label{eq:DOS1reso}
D(\omega)=D_0\frac{\Delta \omega ^2}{(\omega-\omega_0)^2+\Delta \omega^2},
\end{equation}
with the resonance frequency $\omega_0$ and the line width $\Delta \omega$ defined by $Q=\omega_0/\Delta \omega=100$.

\begin{figure}
\includegraphics[width=0.9\linewidth]{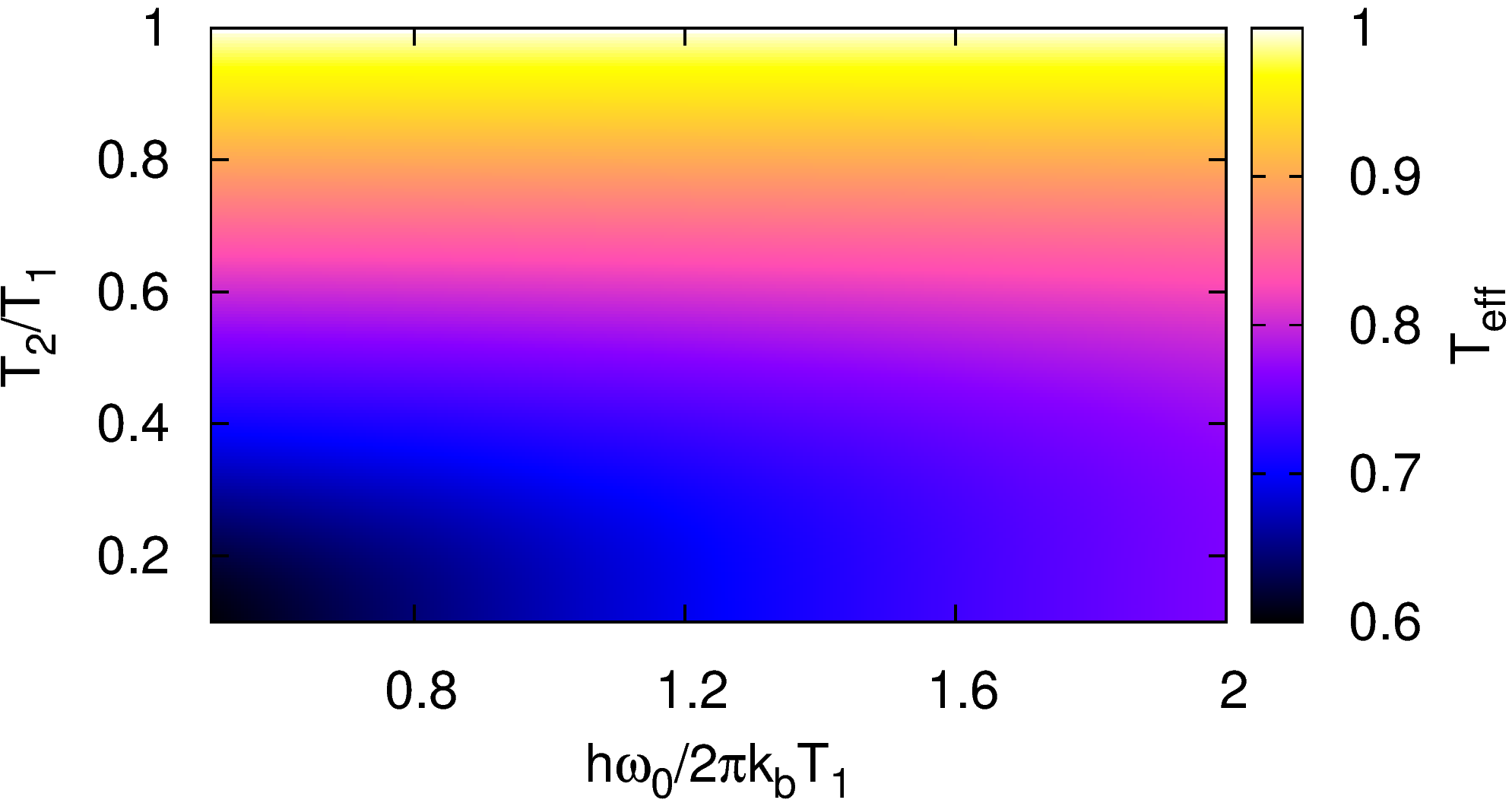}
\includegraphics[width=0.9\linewidth]{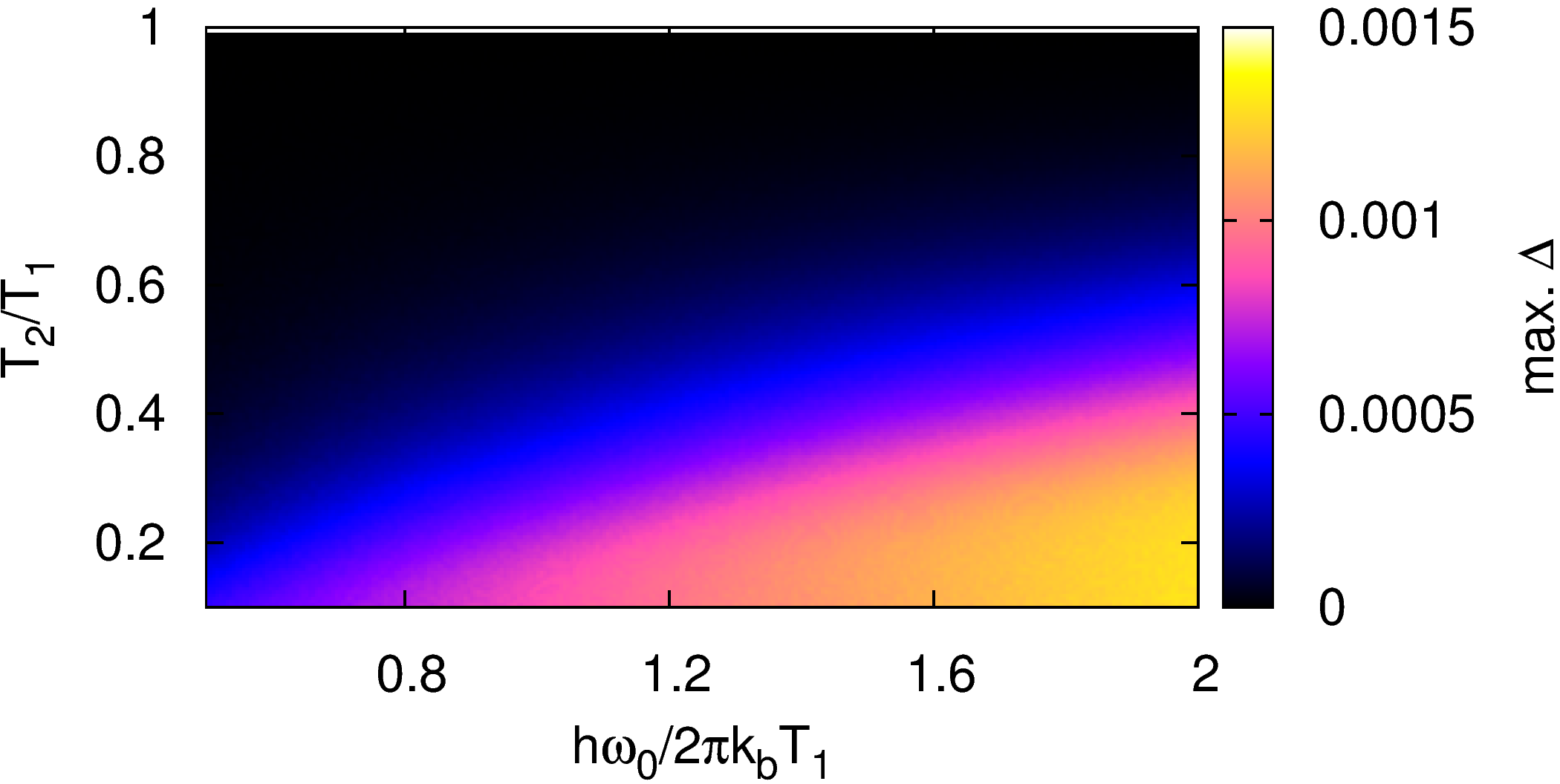}
\caption{\label{fig:1reso} (Color online) Effective temperature (upper panel) and
  maximum difference between the NEST- and equilibrium distribution
  (lower panel) for a cavity with one resonance located at
  $\omega_0$. Due to the single, sharp resonance the description via
  an effective temperature works quite well.}
\end{figure}

The effective temperature and the maximum derivation of the NEST
distribution from an equilibrium one are shown in
Fig. \ref{fig:1reso}. As the density of the state has a sharp peak
around $\omega_0$ the dominant contribution to the integral in
Eq.~(\ref{eq:delta}) steams from frequencies around $\omega_0$. Hence
the description via an equilibrium distribution is quite good, as
demonstrated by the very small errors in Fig. \ref{fig:1reso}.

The situation becomes more interesting by assuming that the cavity has two resonances, so that the DOS is given by
\begin{equation}
\label{eq:DOS2reso}
D(\omega)=D_0 \left(\frac{\Delta \omega_1 ^2}{(\omega-\omega_1)^2+\Delta \omega_1^2}+\frac{\Delta \omega_2 ^2}{(\omega-\omega_2)^2+\Delta \omega_2^2}\right).
\end{equation}

\begin{figure}
\includegraphics[width=0.9\linewidth]{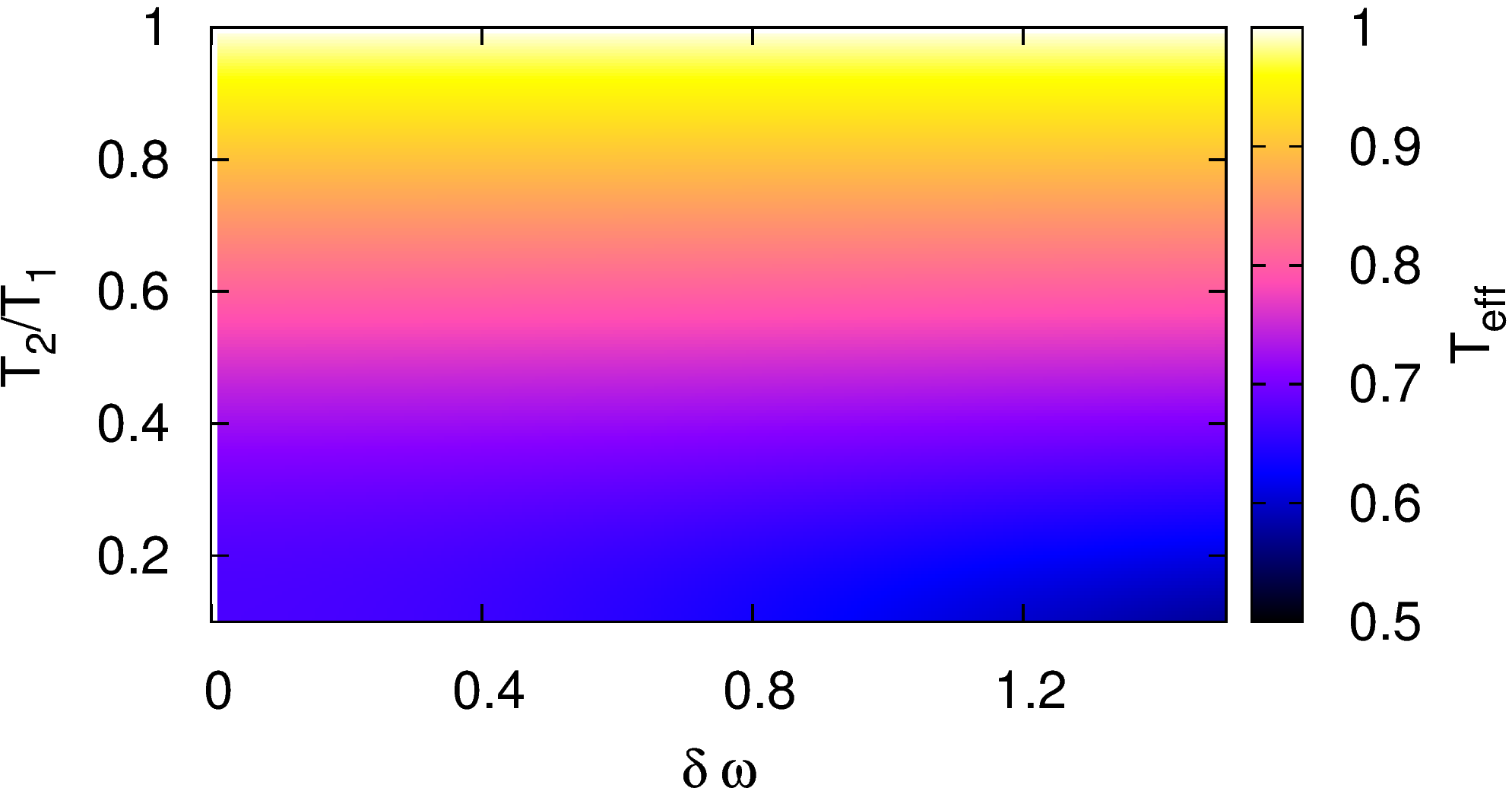}
\includegraphics[width=0.9\linewidth]{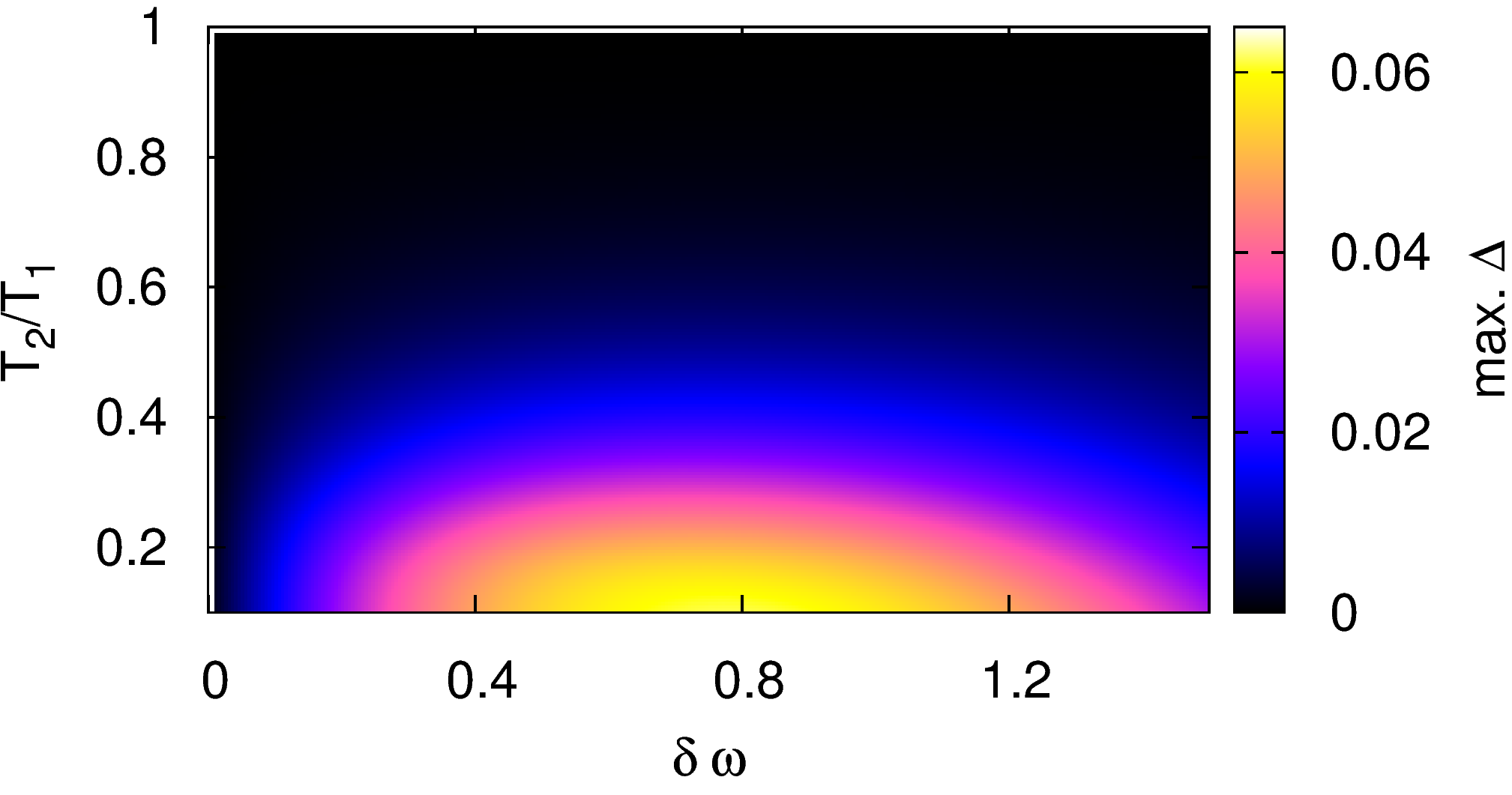}
\caption{\label{fig:2reso} (Color online) Effective temperature (upper panel) and  maximum difference between the NEST- and equilibrium distribution (lower panel) for a cavity with two resonances $\omega_1=(0.8-0.5\delta\omega)\frac{k_b T_1}{\hbar}$ and $\omega_2=(0.8+0.5\delta\omega)\frac{k_b T_1}{\hbar}$.}
\end{figure}

In this case the differences between the NEST and the effective
description are much more pronounced (see Fig.~\ref{fig:2reso}). And
hence the results, presented in this section, provide a guideline,
whether a description via an effective temperature is suitable or
not. If the DOS is dominated by single resonance the system can be
described by an effective temperature, while non-equilibrium effects
have to be included when a broader frequency range comes into play.

\section{\label{sec:conclusion}Conclusion}

We have investigated the non-equilibrium steady state (NEST) for
photons in a cavity for which the boundaries are at two or more
distinct temperatures. For this purpose we have derived an analytic
expression for the NEST photon statistic and evaluated this expression
for two different situations. Once assuming that the density of states
can be described by the vacuum density and secondly for multi-mode
cavities.

\begin{enumerate}
\item If the density of states can be described by vacuum density, we
  predict that the NEST-equivalent of Planck's law can show deviations
  of more than 10\% from the usual form of Planck's law if we take two
  distinct temperatures at the boundaries of the cavity that differ by
  a factor of two. 

\item While for a single-mode cavity the description of the photonen
  statistics via an effective temperature captures the NEST quite
  well, for multi-mode cavities the photon statistics is dominated by
  the resonances and non-equilibrium effects have to be accounted for.

\end{enumerate}

The approach presented here should have practical consequences for the
energy and momentum transfer between bodies of comparable size in a
NEST situation. Furthermore, black-body spectra influence the accuracy of
state-of-the-art atomic clocks~\cite{MiddelmannEtAl2012,BloomEtAl2014,HinkleyEtAl2013}
for which temperature gradients near the atoms have been
reported~\cite{BloomEtAl2014}. Thus, even in the absence of cavity
resonances, calculating deviations from the black-body spectrum caused
by more than one temperature can become important for high-precision
atomic clocks.

For our approach to be valid, the time-scale on which the NEST is
reached has to be shorter than the time-scales at which the
temperatures at the boundaries of the cavity change.

\acknowledgments

We thank S.\ A.\ Biehs, E.\ M.\ Bridge, M.\ Holthaus, I.\ G.\ Hughes, C.\ Lienau, A.\ Kittel and V.\ Steenhoff for discussions.

\begin{appendix}
\section{\label{app:lindblad}Lindblad Master equation}

A derivation of the Lindblad Master equation can be found, e.g., in Ref.~\cite{Breuer06}. If $a^{(\dag)}$ is the annihilation (creation) operator of a photon in mode~$\omega$, the Lindblad equation for the cavity field reads:
\begin{eqnarray}
\label{eq:lindblad}
\frac{d}{dt}\hat{\varrho} =&&
-i\omega [a^{\dag}a,\hat{\varrho}] -\frac{\kappa_-}2\left(a^{\dag}a\hat{\varrho}+\hat{\varrho} a^{\dag}a - 2a\hat{\varrho} a^{\dag}\right)\nonumber\\
&&-\frac{\kappa_+}2\left(aa^{\dag}\hat{\varrho}+\hat{\varrho}a a^{\dag} - 2a^{\dag}\hat{\varrho} a\right)\;,
\end{eqnarray}

where $\hat{\varrho}$ is the density matrix.
The rate equation~(\ref{eq:masterp}) can be derived~\cite{HarocheRaimond06} by using the fact that the photon number distribution $p_n(t)$ is related to the density matrix via
\begin{equation}
%\nonumber
p_n(t) = \langle n|\varrho|n\rangle \;.
\end{equation}
In this paper, the Lindblad Master equation is used to describe
emission and absorption of thermal photons, other possible
applications include atom losses~\cite{StieblerEtAl2011}.
\end{appendix}
%\newpage
\bibliography{photonen_stat}{}

\end{document}